\documentclass[aps,english,prb,twocolumn,reprint,superscriptaddress]{revtex4}
\usepackage{graphicx}
\usepackage[left]{lineno}
\usepackage{gensymb}
\usepackage{amsmath}

\makeatletter
\def\p@figure{}
\makeatother

\usepackage[version=4]{mhchem}

\begin{document}

\title{Composition and Surface Termination Control Band Alignments in P3HT/Perovskite Heterostructures}

\author{Somayyeh Alidoust}
\affiliation{Faculty of Engineering and Natural Sciences, Sabanci University, Istanbul, 34956 Turkey.}

\author{V. Ongun {\"O}z{\c{c}}elik} \email{ongun.ozcelik@sabanciuniv.edu}
\affiliation{Faculty of Engineering and Natural Sciences, Sabanci University, Istanbul, 34956 Turkey.}
\affiliation{Materials Science and Nano Engineering Program, Sabanci University, Istanbul, 34956 Turkey.}

\begin{abstract}
		Hybrid organic–inorganic perovskite heterostructures are emerging platforms for energy conversion and optoelectronic devices, yet the effects of composition and surface terminations on band alignment remain poorly understood. Here, using high-throughput first-principles calculations, we systematically reveal the effect of composition and surface termination on the band alignment of 820 heterostructures formed between poly(3-hexylthiophene) (P3HT) variants and conventional and mixed-cation metal halide perovskites. Band edge analysis predicts concentration dependent band alignments, with most Pb and Sn based perovskites exhibiting Type II and III behavior while mixed-cation CsMASnBr$_3$ uniquely forms Type-I junctions with all P3HT variants. Explicit interface calculations reveal that surface termination can reverse the predicted alignment in conventional tin perovskites, while the mixed cation perovskite CsMASnBr$_3$ preserves Type-I alignment with P3HT regardless of termination. Despite negligible charge transfer, termination-dependent band offsets produce markedly different carrier confinement, establishing design principles for engineering P3HT/perovskite interfaces.
\end{abstract}

\maketitle

\newpage

Organic–inorganic hybrid heterostructures are central to next generation energy conversion technologies, where metal halide perovskites have emerged as leading materials for high-efficiency photovoltaics. \cite{zhang2016metal,stranks2015metal,wang2021prospects,tong2021wide,dai2022pathways} Their interfacial electronic structure and band alignment type governs charge separation, carrier recombination, and energy losses, which determine device performance.\cite{chen2023universal,10.1021/acs.jpcc.9b10055,mater2026interfacial} Metal halide perovskites possess outstanding optoelectronic properties, including strong optical absorption, tunable band gaps, long carrier diffusion lengths, and high charge-carrier mobility, making them highly attractive absorber materials for next-generation solar cells \cite{jiang2024rapid}{.} However, efficient charge extraction critically depends on achieving favorable band alignment with adjacent organic semiconductors \cite{10.1021/acsaem.5c04025, https://doi.org/10.1002/anie.202313574}{.} Consequently, rational interface engineering through band alignment engineering and surface treatement has become an essential strategy for improving device performance and stability.\cite{chen2020materials,ji2026critical} Among the organics used in these hybrid materials, poly(3-hexylthiophene) (P3HT) is an attractive organic semiconductor owing to its high hole mobility, chemical stability, solution processability, and compatibility with low-cost device fabrication.\cite{fratini2020charge,sirringhaus1999,marrocchi2012p3ht} P3HT has been extensively employed as an active semiconductor or hole-transport material in hybrid perovskite devices.\cite{https://doi.org/10.1002/cssc.202500460} While numerous experimental and theoretical studies have examined P3HT and its interfaces with inorganic materials, \cite{10.1021/nl801700s,10.1021/acs.jpcc.0c03543,mater2025interfacialchargetransferelectronic,10.1021/acs.jpclett.2c02130,Xiang2017,Dag2008, Adeniran2020} these studies demonstrate that interfacial electronic structure and charge-transfer behavior can depend strongly on the specific composition and interface, making it difficult to establish general design principles for optimal material combinations.

A major unresolved challenge is understanding how perovskite composition and interface structure collectively determine band alignment. Band edge positions obtained from isolated materials provide a convenient data base, yet interface formation can introduce structural relaxation, orbital hybridization, and surface-dependent effects that substantially modify the electronic structure. Similarly, it remains unclear whether band alignments predicted from isolated components are preserved after realistic interface formation or whether they change into different heterojunction types. Moreover, the influence of mixed A-site cations on these interfacial electronic properties has received little systematic attention despite their widespread use in high-performance perovskite materials.

In this letter, we perform a comprehensive first-principles investigation of band alignment in heterostructures formed between multiple structural variants of P3HT and a broad family of conventional ABX$_3$ and mixed-cation AA$'$BX$_3$ metal halide perovskites, where A/A$'$ denote monovalent organic or inorganic cations (e.g., methylammonium (MA), formamidinium (FA), and Cs), B is a divalent metal cation (e.g., Pb or Sn), and X is a halide anion. Vacuum-referenced band-edge analysis establishes composition-dependent trends across the materials space and predicts Type-I, Type-II, and Type-III heterojunctions based on the relative band-edge positions.\cite{PhysRevB.94.035125} Our explicit interface calculations further reveal that surface termination can substantially alter the predicted band alignment in conventional tin perovskites, whereas the mixed-cation perovskite CsMASnBr$_3$ uniquely preserves Type-I alignment regardless of termination. The calculated band offsets demonstrate highly tunable carrier confinement with negligible charge transfer, indicating that orbital coupling rather than charge redistribution governs the interfacial electronic structure. These results establish practical design rules for engineering P3HT/perovskite heterostructures and identify mixed-cation chemistry as an effective route for tailoring electronic interfaces in hybrid energy materials.

\begin{figure*}[htbp]
	\centering
	\includegraphics[ width=1\textwidth]{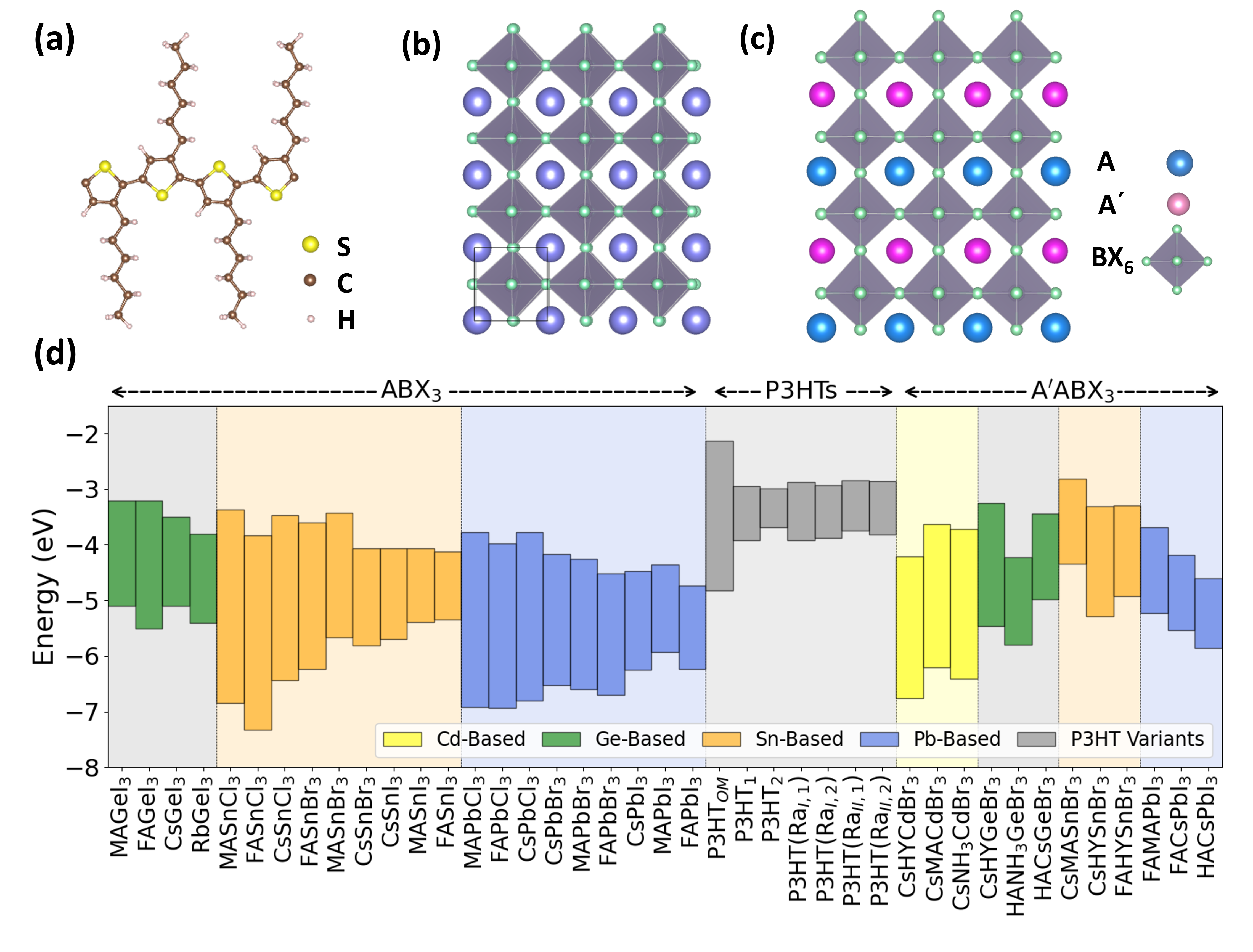}
	\caption{(a) Crystal structure of the P3HT polymer. (b) Schematic of the mixed-cation AA$'$BX$_3$ perovskite. (c) Band-edge positions of the P3HT variants compared with conventional ABX$_3$ perovskites from the literature \cite{10.1116/6.0001903,https://doi.org/10.1002/pssa.202200791} and the mixed-cation AA$'$BX$_3$ perovskites calculated in this study. I and II denote two random alkyl-chain arrangements, while subscripts 1 and 2 indicate monolayer and bilayer P3HT, respectively.}
	\label{fig:fig1}
\end{figure*}

We start our analyses by considering seven P3HT variants, 22 conventional ABX$_3$ perovskites \cite{10.1116/6.0001903,https://doi.org/10.1002/pssa.202200791} and 12 mixed A-site AA$'$BX$_3$ perovskite structures selected from a pool of promising compositions reported in our previous DFT studies. \cite{Alidoust2024,D4CP04218B} The crystal structure of P3HT and representative crystal structures of conventional ABX$_3$ and mixed-cation AA$'$BX$_3$ perovskites are presented in Figure~\ref{fig:fig1}. Subsequently, DFT calculations were performed on slab models of the P3HT variants and mixed-cation perovskites to evaluate their VBM and CBM positions, while the band-edge positions of conventional ABX$_3$ perovskites were taken from the literature. \cite{10.1116/6.0001903,https://doi.org/10.1002/pssa.202200791}\\
All first-principles calculations were carried out using the Vienna Ab initio Simulation Package.\cite{PhysRevB.54.11169} The electronic structure was treated within the generalized gradient approximation  using the Perdew–Burke–Ernzerhof\cite{PhysRevLett.77.3865} exchange–correlation functional. Long-range van der Waals interactions were included via the DFT-D3 correction\cite{10.1063/1.5023802}{.} Electron–ion interactions were described using the projector-augmented-wave  metho.\cite{PhysRevB.50.17953}

For each system, full geometry optimizations were performed by relaxing all atomic positions and cell volumes while keeping the lattice angles fixed to maintain the overall cell shape. The plane-wave basis set was expanded up to a cutoff energy of 500 eV, and a k-point mesh of approximately 40 points along each periodic direction was employed to ensure convergence. Electr onic self-consistency was achieved with an energy criterion of 10$^{-6}$ eV, and ionic relaxations were continued until the residual forces on all atoms were below 10$^{-5}$ eV/Å.

Subsequent electronic structure calculations, including band gap determination, were carried out using the screened hybrid functional (HSE06) with spin–orbit coupling (SOC). A fraction of 25\% exact Hartree–Fock exchange was included in the hybrid functional to improve the accuracy of band edges and gap values. For slab and interface models, a vacuum layer of 30 Å was applied along the non-periodic direction to avoid spurious interactions between periodic images, and dipole corrections were consistently applied to eliminate artificial electrostatic effects.

Band alignment at the perovskite/polymer interfaces was determined by referencing all electronic energy levels to a common vacuum level, obtained from the planar-averaged electrostatic potential along the surface-normal direction where a flat potential plateau was observed. The VBM, CBM, and Fermi level of each system were then aligned relative to this vacuum reference.

Figure ~\ref{fig:fig1}d compares the band-edge positions of the conventional ABX$_3$ perovskites, P3HT variants, and mixed-cation AA$'$BX$_3$ perovskites. Ge-, Sn-, and Pb-based ABX$_3$ perovskites exhibit similar electronic trends, with iodides showing narrower band gaps than the corresponding bromides and chlorides, placing their band edges closer to the visible-light absorption region. In contrast, the electronic energy levels of the P3HT variants depend on the polymer structure, reflecting the effects of chain conformation and interchain packing. Although these thin-layer models do not fully represent bulk P3HT, they provide useful insight into how structural ordering influences the electronic structure, charge confinement, and charge-transfer characteristics. The mixed-cation AA$'$BX$_3$ perovskites generally follow the same trends as their conventional counterparts, while Cd-based compounds stand out with deeper valence bands, wider band gaps, and larger energy offsets than the Ge-, Sn-, and Pb-based systems. Table~S1 of the Supporting Information (SI)summarizes the calculated work functions ($\Phi$) of the mixed-cation perovskites and P3HT variants, which characterize the energetic position of the Fermi level relative to the vacuum level and are relevant to charge transfer and the energetic conditions governing charge confinement and energy-level alignment. The relatively broad range across both material classes (3.5–6.1 eV) highlights the potential for tuning interfacial energy-level alignment through compositional and structural modifications. Overall, both halide composition and B-site chemistry primarily determine the absolute band-edge positions, giving rise to Type-I, Type-II, and Type-III band alignments with P3HT.

\begin{figure*}[htbp]
	\centering
	\includegraphics[width=1.\textwidth, angle=0]{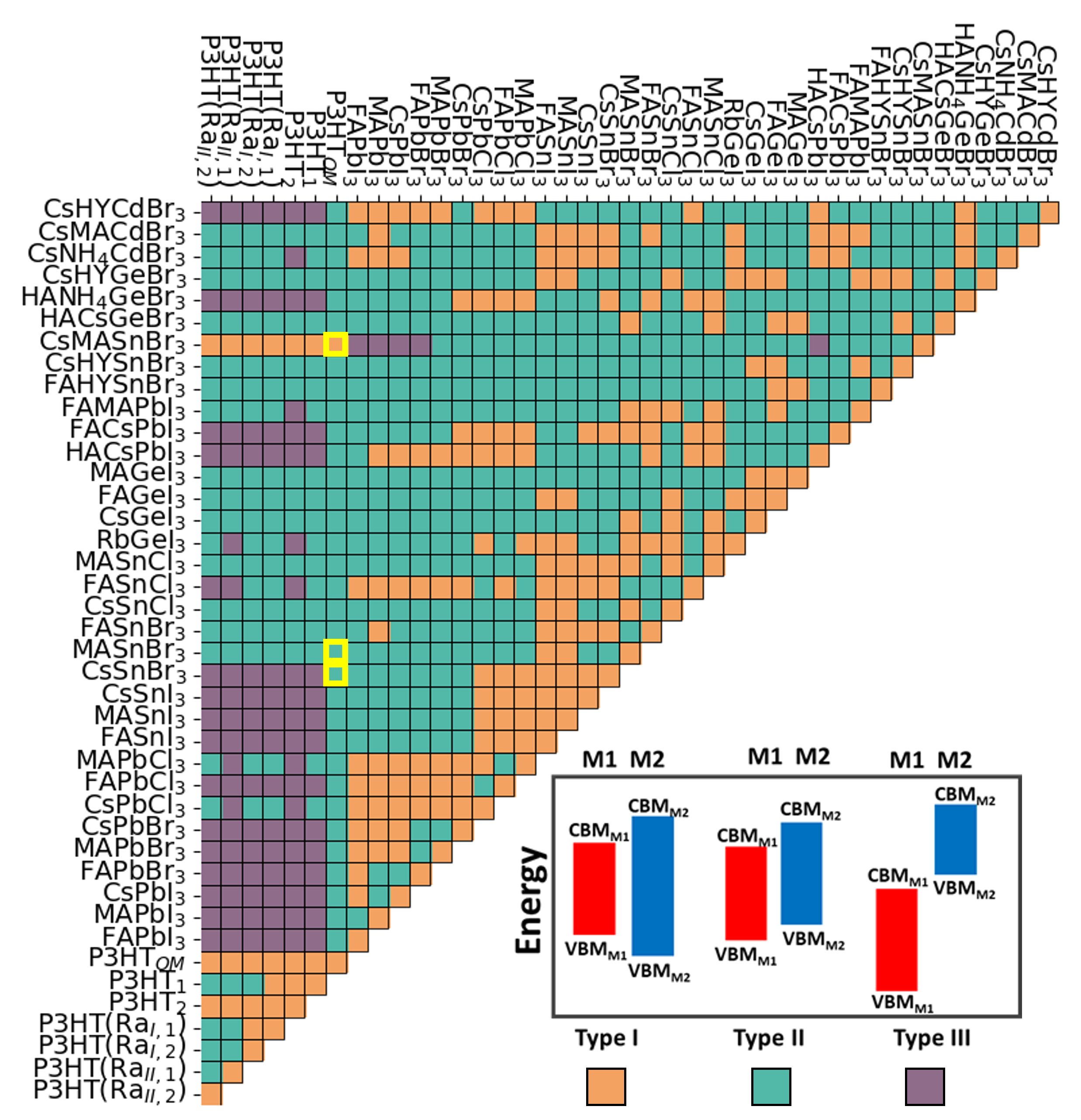}
	\vspace{-5pt}
	\caption{Band alignment heat map for perovskites and P3HT variants. Yellow squares mark selected Sn-based perovskites (CsSnBr$_3$, MASnBr$_3$, CsMASnBr$_3$) for interface modeling.}
	\label{fig:fig2}
\end{figure*}

A heatmap summarizing the band-alignment types between the investigated perovskites and different P3HT configurations, comprising a total of 820 heterostructures, is presented in Figure~\ref{fig:fig2}. The VBM and CBM energies of the perovskites and P3HT variants are also presented in Table \ref{tab:vbm_cbm}.Overall, most perovskite/P3HT heterostructures exhibit either Type-II (teal) or Type-III (muted plum) band alignment, indicating that staggered and broken-gap junctions are generally favored in these systems. Pb-based perovskites predominantly form Type-II alignments with P3HT, while SnI-based perovskites also consistently exhibit Type-II behavior. In contrast, SnBr-based and Cd-based perovskites display a mixture of Type-II and Type-III alignments depending on the perovskite composition and P3HT configuration. A notable exception is the mixed-cation CsMASnBr$_3$, which forms a Type-I (orange) alignment with all considered P3HT structures, distinguishing it from the other investigated materials. In contrast, the corresponding pure compounds, CsSnBr$_3$ and MASnBr$3$, predominantly exhibit Type-III and Type-II alignments with P3HT, respectively, although both form Type-I alignments with P3HT${\mathrm{OM}}$. This striking difference highlights the significant influence of cation mixing on interfacial band alignment. To validate these predictions, we explicitly modeled the interfaces between three representative less-toxic tin-based perovskites—CsSnBr$_3$, MASnBr$_3$, and CsMASnBr$3$—and  a hydrogen-passivated oligomeric P3HT configuration (P3HT$_{OM}$), as highlighted by the yellow rectangles in Figure \ref{fig:fig2}. This comparison reveals the effect of A-site cation mixing on the interfacial electronic properties, particularly charge transfer and carrier confinement.

\begin{table*}[htbp]
	\centering
	\caption{Valence band maximum (VBM) and conduction band minimum (CBM) energies (eV) referenced to the vacuum level for perovskites and P3HT variants.}
	\label{tab:vbm_cbm}
	\small
	\begin{tabular}{|lcc|lcc|}
		\hline
		\hline
		\multicolumn{3}{|c|}{\rule{0pt}{2.8ex}\textbf{AA$'$BX$_3$ Perovskites}} & 
		\multicolumn{3}{c|}{\rule{0pt}{2.8ex}\textbf{ABX$_3$ Perovskites}} \\
		\hline
		\rule{0pt}{2.8ex}Material & CBM/k-point & VBM/k-point & 
		Material & CBM & VBM \\
		\hline
		
		CsHYCdBr$_3$      & $-4.22$/$\Gamma$ & $-6.77$/Y & FAGeI$_3$     & $-3.20$ & $-5.50$ \\
		CsMACdBr$_3$      & $-3.63$/X & $-6.21$/X & CsGeI$_3$     & $-3.50$ & $-5.10$ \\
		CsNH$_3$CdBr$_3$  & $-3.72$/X & $-6.41$/X & RbGeI$_3$     & $-3.80$ & $-5.40$ \\
		CsHYGeBr$_3$      & $-3.25$/X & $-5.46$/$\Gamma$ & MASnCl$_3$    & $-3.36$ & $-5.85$ \\
		HANH$_3$GeBr$_3$  & $-4.23$/C & $-5.80$/C & FASnCl$_3$    & $-3.83$ & $-7.33$ \\
		HACsGeBr$_3$      & $-3.44$/Y & $-4.98$/$\Gamma$ & CsSnCl$_3$    & $-3.47$ & $-5.44$ \\
		CsMASnBr$_3$      & $-2.81$/Y & $-4.35$/Y & FASnBr$_3$    & $-3.60$ & $-6.23$ \\
		CsHYSnBr$_3$      & $-3.31$/X & $-5.30$/X & MASnBr$_3$    & $-3.42$ & $-5.67$ \\
		FAHYSnBr$_3$      & $-3.29$/X & $-4.93$/X & CsSnBr$_3$    & $-4.07$ & $-5.82$ \\
		FAMAPbI$_3$       & $-3.69$/Y & $-5.23$/Y & CsSnI$_3$     & $-4.07$ & $-5.69$ \\
		FACsPbI$_3$       & $-4.19$/Y & $-5.54$/Y & MASnI$_3$     & $-4.07$ & $-5.39$ \\
		HACsPbI$_3$       & $-4.61$/Y & $-5.85$/Y & FASnI$_3$     & $-4.12$ & $-5.34$ \\
		\cline{1-3}
		& \rule{0pt}{2.8ex}\textbf{P3HTs} &  & MAPbCl$_3$    & $-3.77$ & $-6.92$ \\		
		\cline{1-3}
		\rule{0pt}{2.8ex} Material & CBM/k-poin & VBM/k-poin & FAPbCl$_3$  & $-3.98$ & $-6.94$\\
		\cline{1-3}
		\rule{0pt}{2.8ex}P3HT$_{OM}$        & $-2.12$/$\Gamma$ & $-4.83$/$\Gamma$ &  CsPbCl$_3$ &$-3.77$ & $-6.80$ \\
		P3HT$_1$        & $-2.94$/$\Gamma$ & $-3.92$/$\Gamma$ &   CsPbBr$_3$    & $-4.17$ & $-6.53$ \\
		P3HT$_2$        & $-2.99$/$\Gamma$ & $-3.68$/$\Gamma$ &    MAPbBr$_3$    & $-4.25$ & $-6.60$\\
		P3HT(Ra$_{I,1}$) & $-2.87$/$\Gamma$ & $-3.92$/$\Gamma$ &  FAPbBr$_3$    & $-4.51$ & $-6.70$\\
		P3HT(Ra$_{I,2}$) & $-2.93$/$\Gamma$ & $-3.87$/$\Gamma$ &  CsPbI$_3$     & $-4.47$ & $-6.25$ \\
		P3HT(Ra$_{II,1}$) & $-2.84$/$\Gamma$ & $-3.75$/$\Gamma$ & MAPbI$_3$    & $-4.36$ & $-5.93$ \\
		P3HT(Ra$_{II,2}$) & $-2.85$/$\Gamma$ & $-3.81$/$\Gamma$ &  FAPbI$_3$  & $-4.74$ & $-6.24$\\
		&         &         & MAGeI$_3$     & $-3.20$ & $-5.10$ \\
		
		\hline
	\end{tabular}
\end{table*}

By performing explicit DFT interface calculations, we aim to systematically examine whether the band alignment characteristics predicted from isolated materials are preserved or modified upon interface formation. This approach enables a consistent comparison of their interfacial electronic structures and band alignment types, providing deeper insight into the interplay between composition and interfacial charge behavior, including possible charge redistribution and carrier confinement effects.\\
In order to determine the optimal interface configurations, various perovskite/P3HT$_{OM}$ heterostructures were constructed considering different surface terminations (AX and BX) of CsSnBr$_3$ and MASnBr$_3$. Previous experimental studies suggest that the surface termination of solution-processed MAPbI$_3$ films is influenced by the local chemical environment and surface reconstruction effects, rather than being uniquely determined by a thermodynamic ground state. \cite{Mirzehmet2021} Therefore, both AX- and BX-terminated perovskite surfaces, which are likely to coexist under solution-processing conditions and have been widely considered in previous studies,\cite{KasiMatta2018} were included in the interface modeling. Figures S2–S3 in the SI present the different interface configurations together with their corresponding adsorption energies for the selected perovskites. The most stable configurations were subsequently selected for further electronic structure analysis.
\begin{figure*}[htbp]
	\centering
	\includegraphics[width=\textwidth]{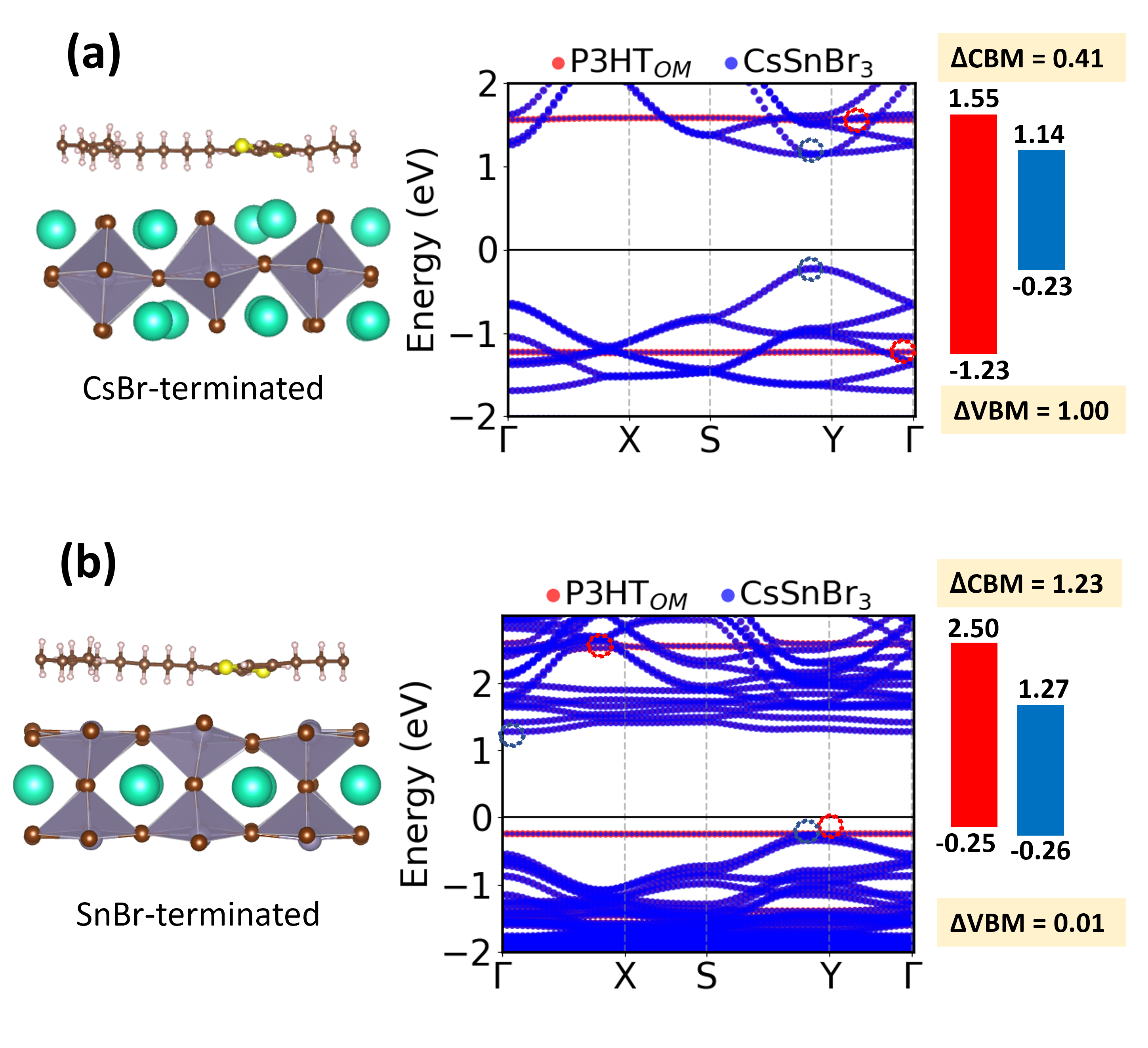}
	\caption{Optimized interfacial atomic structures, projected band structures, and band alignments of the CsSnBr$_3$/P3HT$_{OM}$ heterostructure for (a) CsBr-terminated and (b) SnBr-terminated interfaces. Dashed circles highlight the corresponding VBM and CBM positions of CsSnBr$_3$ and P3HT$_{OM}$.}
	\label{fig:fig3}
\end{figure*}

\begin{figure*}[htbp]
	\centering
	\includegraphics[width=\textwidth]{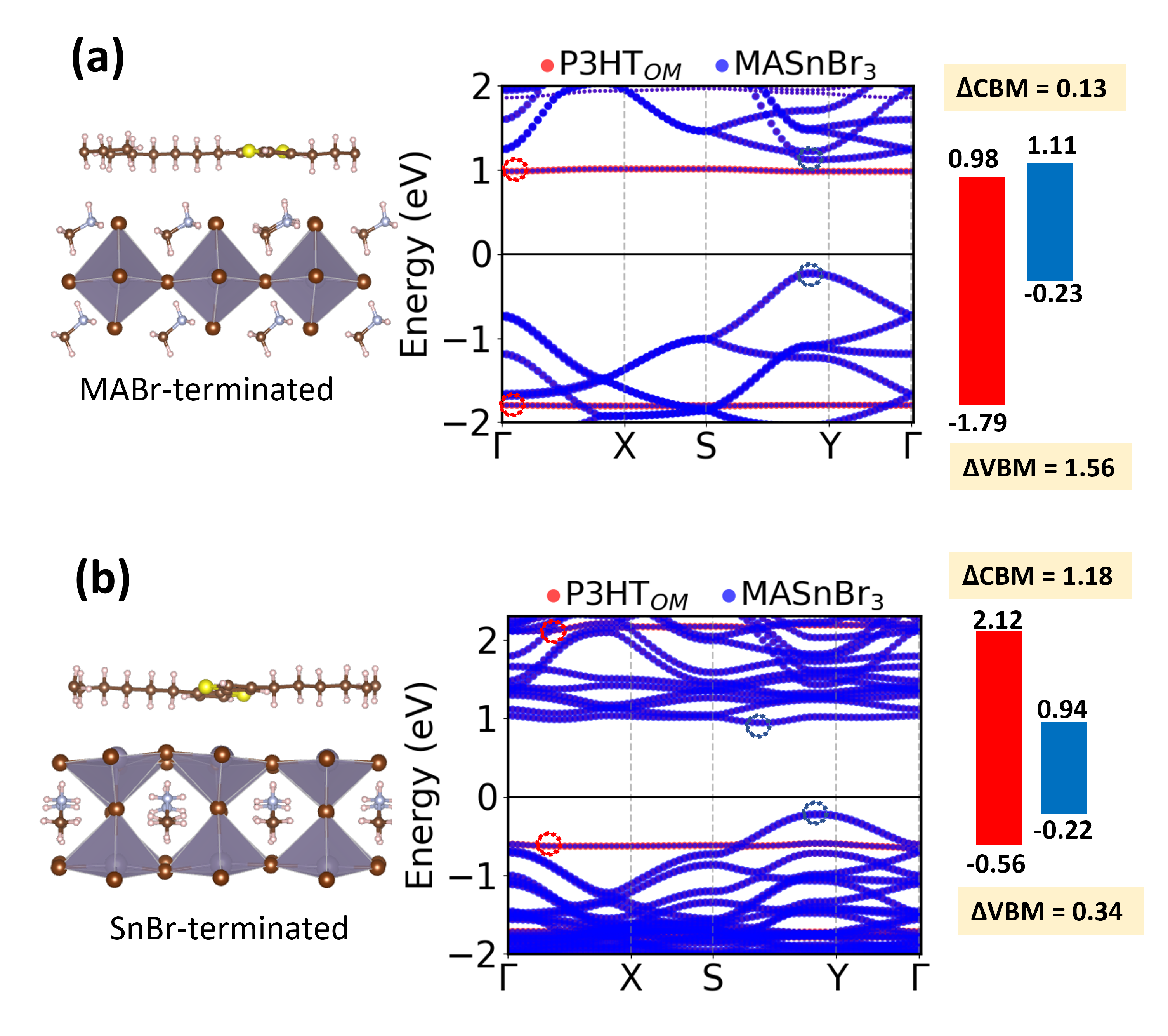}
	\caption{Optimized interfacial atomic structures, projected band structures, and band alignments of the MASnBr$_3$/P3HT$_{OM}$ heterostructure for (a) MABr-terminated and (b) SnBr-terminated interfaces. Dashed circles highlight the corresponding VBM and CBM positions of MASnBr$_3$ and P3HT$_{OM}$.}
	\label{fig:fig4}
\end{figure*}

\begin{figure*}[htbp]
	\centering
	\includegraphics[width=\textwidth]{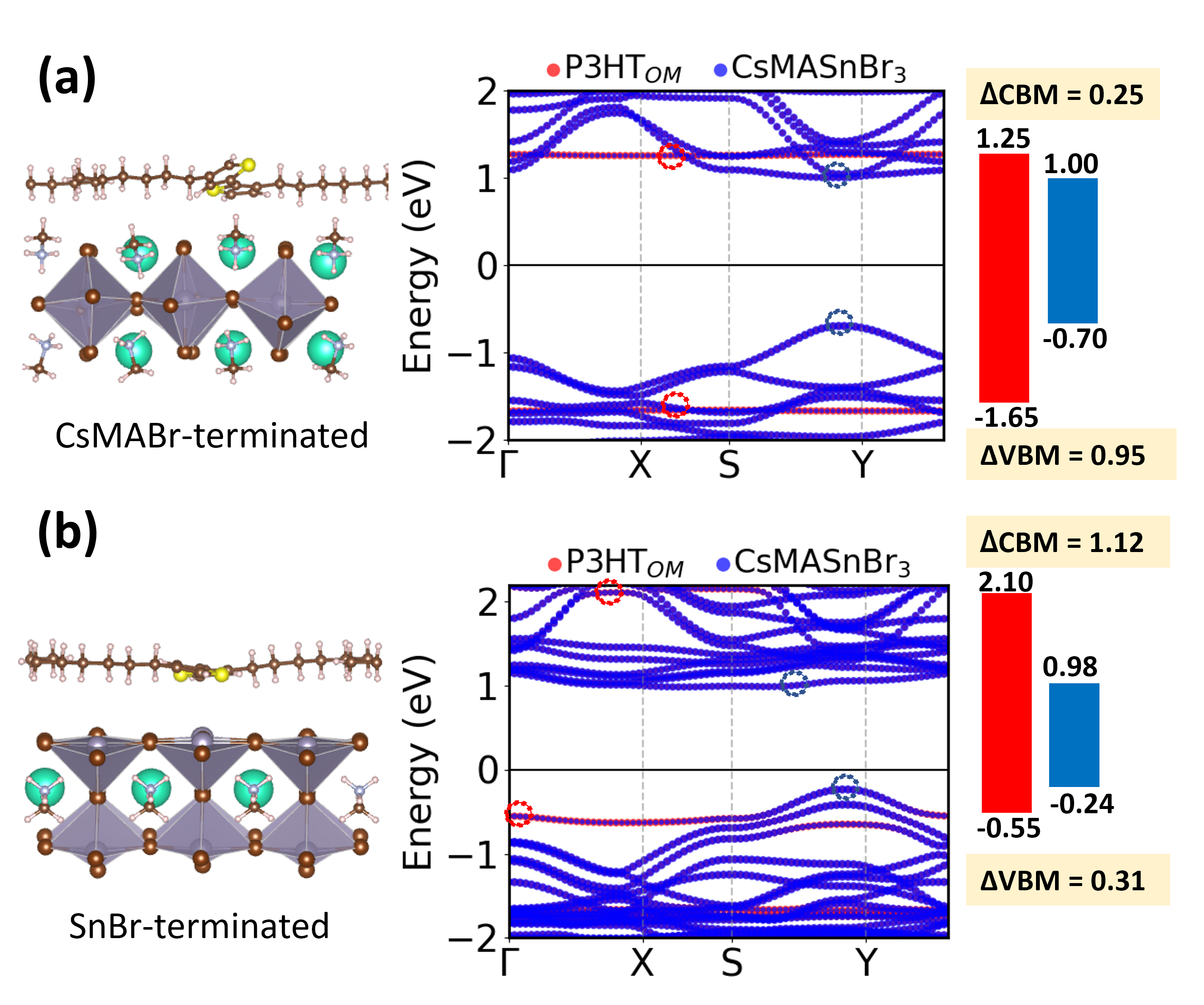}
	\caption{Optimized interfacial atomic structures, projected band structures, and band alignments of the CsMASnBr$_3$/P3HT$_{OM}$ heterostructure for (a) CsMABr-terminated and (b) SnBr-terminated interfaces. Dashed circles highlight the corresponding VBM and CBM positions of CsMASnBr$_3$ and P3HT$_{OM}$.}
	\label{fig:fig5}
\end{figure*}

Figure \ref{fig:fig3} and \ref{fig:fig4} present the optimized crystal structures and projected band structures of the CsSnBr$_3$/P3HT$_{OM}$ and MASnBr$_3$/P3HT$_{OM}$ heterostructures, with electronic states projected onto the perovskite and P3HT$_{OM}$ components. The results reveal that the type-II band alignment predicted for isolated CsSnBr$_3$ transforms into a type-I alignment in the heterostructure with AX (CsBr)-terminated surfaces, while it remains type II for the BX (SnBr)-terminated interface. In the case of MASnBr$_3$/P3HT$_{OM}$, the type-II alignment of the isolated materials is preserved for the AX (MABr)-terminated interface but changes to type I for the BX (SnBr)-terminated heterostructure. The calculated CBM and VBM offsets between CsSnBr$_3$/P3HT$_{OM}$ depend on the surface termination. For the CsBr-terminated interface, the offsets are $\Delta CBM = 0.41$ eV and $\Delta VBM = 1.0$ eV, indicating relatively weak electron confinement but strong hole confinement within CsSnBr$_3$. In contrast, the SnBr-terminated interface exhibits $\Delta CBM = 1.23$ eV and $\Delta VBM = 0.01$ eV, suggesting strong electron confinement and negligible hole confinement within CsSnBr$_3$. Similar to the CsSnBr$_3$/P3HT$_{OM}$ heterostructure, the band offsets in MASnBr$_3$/P3HT$_{OM}$ are highly sensitive to the surface termination. The MABr-terminated interface exhibits $\Delta CBM = 0.13$ eV and $\Delta VBM = 1.56$ eV, corresponding to weak electron confinement and strong hole confinement. In contrast, the SnBr-terminated interface yields $\Delta CBM = 1.18$ eV and $\Delta VBM = 0.34$ eV, indicating strong electron confinement together with moderate hole confinement.\\
The optimized configurations of CsSnBr$_3$/P3HT$_{OM}$ and MASnBr$_3$/P3HT$_{OM}$ heterostructures were then used as guides to construct and refine the interface of the mixed perovskite CsMASnBr$_3$/P3HT$_{OM}$, enabling the preparation of a more stable and physically reasonable configuration for subsequent analysis. Figure \ref{fig:fig5} shows the optimized crystal structures of the CsMASnBr$_3$/P3HT$_{OM}$ heterostructures for AX(CsMABr)- and BX(SnBr)-terminated perovskite slabs, respectively, along with their corresponding projected band structures. The type I alignment observed in the individual materials (CsMASnBr$_3$ and P3HT$_{OM}$) is preserved in the heterostructure for both surface terminations, CsMABr and SnBr, indicating that the relative positions of the valence and conduction band edges remain largely unchanged upon interface formation. 
Similar to the corresponding pure perovskites, the calculated CBM and VBM offsets between CsMASnBr$_3$ and P3HT$_{OM}$ are highly sensitive to the surface termination. The CsMABr-terminated interface exhibits $\Delta CBM = 0.25$ eV and $\Delta VBM = 0.95$ eV, corresponding to weak electron confinement and strong hole confinement. In contrast, the SnBr-terminated interface yields $\Delta CBM = 1.12$ eV and $\Delta VBM = 0.31$ eV, indicating strong electron confinement and moderate hole confinement. Consistent with the behavior of the pure CsSnBr$_3$ and MASnBr$_3$ compounds, these results highlight the important role of surface termination in determining carrier confinement at the perovskite/P3HT interface. The resulting asymmetry in the band offsets may influence carrier localization and recombination dynamics. \\
We further present the Bader charge transfer for the three heterostructures considering both surface terminations of the perovskites. The results, summarized in Table S2, reveal that the net charge transfer is minimal for all interfaces (approximately 0.01--0.03~e). Despite this negligible charge transfer, significant differences are observed in both the magnitude and nature of the band gap between the two terminations. For example, in the CsMASnBr$_3$/P3HT$_{OM}$ heterostructure, the CsMABr-terminated interface exhibits a direct band gap of 1.7 eV, whereas the SnBr-terminated interface shows a smaller indirect band gap of 1.2 eV. This variation indicates that surface termination not only influences the band gap value but also alters the fundamental electronic transition characteristics of the system. This demonstrates that surface termination plays a critical role in modulating the interfacial electronic structure, primarily through differences in orbital overlap and coupling strength rather than through direct charge transfer. While the A-site cations do not directly contribute to the frontier orbitals, they may still influence the electronic properties indirectly via structural effects. Overall, the electronic structure of the heterostructure is governed by carrier confinement, making it potentially suitable for light-emitting and optoelectronic applications rather than photovoltaic charge extraction.\\

In conclusion, we performed a first-principles investigation of band alignment between poly(3-hexylthiophene) (P3HT) variants and metal halide perovskites, including conventional \ce{ABX$_3$} and mixed-cation \ce{AA$'$BX$_3$} compounds. Vacuum-level-referenced band-edge analysis combined with explicit interface calculations reveals how perovskite composition, polymer structure, and surface termination govern interfacial electronic properties. Most Pb- and Sn-based perovskites form Type-II or Type-III heterojunctions with P3HT, whereas the mixed-cation perovskite \ce{CsMASnBr$_3$} consistently exhibits Type-I alignment, highlighting the importance of A-site cation engineering. Explicit interface calculations further show that surface termination can alter the predicted band alignment in \ce{CsSnBr$_3$} and \ce{MASnBr$_3$}, while \ce{CsMASnBr$3$}/P3HT$_{OM}$ preserves Type-I alignment for both terminations. Termination-dependent band offsets produce tunable carrier confinement despite negligible charge transfer, indicating that orbital hybridization and confinement dominate the interfacial electronic behavior. These findings provide practical guidelines for designing P3HT/perovskite heterostructures for optoelectronic applications

\section*{Acknowledgments}
We acknowledges funding from the Scientific and Technological Research Council of Turkey (TUBITAK) under project no 123F264.  Computational resources have been provided by TUBITAK ULAKBIM, High Performance and Grid Computing Center and by the National Center for High Performance Computing of Turkey (UHeM) under grant numbers 1013432022.

\clearpage
\newpage

\bibliographystyle{unsrt}

\bibliography{references.bib}


\end{document}